\documentclass[journal]{IEEEtranTIE}
\usepackage{graphicx}
\usepackage{cite}
\usepackage{picinpar}
\usepackage{amsmath}
\usepackage{url}
\usepackage{flushend}
\usepackage[utf8]{inputenc}
\usepackage{colortbl}
\usepackage{soul}
\usepackage{multirow}
\usepackage{pifont}
\usepackage{textcomp}
\usepackage{color}
\usepackage{alltt}
\usepackage[hidelinks]{hyperref}
\usepackage{enumerate}
\usepackage{siunitx}
\usepackage{breakurl}
\usepackage{epstopdf}
\usepackage{pbox}
\usepackage{caption}
\usepackage{subcaption}
\usepackage{comment}
\usepackage{float}

\begin{document}
\title{
    Conditional Nearest Level Modulation for Improved Switching Dynamics in Asymmetric Multilevel Converters
}

\author{
	\vskip 1em
	Jinshui Zhang, \emph{Student Member},
	Angel V Peterchev, \emph{Senior Member},
	Stefan M Goetz, \emph{Member}
    \thanks{
    This work was supported by Grant No.1RF1MH124943 from the National Institute of Mental Health of the National Institutes of Health of the United States of America; the content is solely the responsibility of the authors and does not necessarily represent the official views of the funding agency.
    }
}

\maketitle

\begin{abstract}
    Modular multilevel converters have promising applications in clean energy, electric vehicles, and biomedical instrumentation, but need many modules to achieve fine output granularity, particularly of the voltage. Asymmetric multilevel circuits introduce differences in module voltages so that the quantity of output levels grows exponentially with the number of modules. Nearest-level modulation (NLM) is preferred over carrier-based methods in asymmetric circuits for its simplicity. However, the large number of output levels can overwhelm NLM and cause excessive transistor switching on some modules and output voltage spikes. We propose a conditional nearest-level modulation (cNLM) by incorporating mathematical penalty models to regulate switching dynamics. This approach improves output quality and reduces switching rates. Additionally, we present cNLM variations tailored for specific functions, such as enforcing a minimum switching interval. Experimental validation on an asymmetric multilevel prototype demonstrates that cNLM reduces the total output distortion from 66.3\% to 15.1\% while cutting the switching rate to just 8\% of the original NLM.
\end{abstract}


{}

\definecolor{limegreen}{rgb}{0.2, 0.8, 0.2}
\definecolor{forestgreen}{rgb}{0.13, 0.55, 0.13}
\definecolor{greenhtml}{rgb}{0.0, 0.5, 0.0}

\section{Introduction}
Multilevel converters, such as cascaded bridge converters and modular multilevel converters (MMCs), have become a key technology in various energy applications---particularly where high quality is desirable and filters are not an option for space reasons or due to their parasitic influence on the load impedance or frequency behavior---such as offshore wind power, high-voltage direct current transmission, and large industrial drives \cite{4591920, 5544594, 5637505, 6038879, 6085335, 5984128, 6397383, 5279054, 8601394, 6342749, 6231806}. The ability to handle simultaneously high power, wide output bandwidth, and high output quality extends MMC's applications beyond conventional energy systems and holds promise for electric vehicle motor drives \cite{li2016hybrid, kumar2016control} as well as pulse generators and biomedical instrumentation \cite{6347016, li_av, zhang2025asymmetric, 10693936, zeng2022modular, sorkhabi2022pulse}.

The fidelity of multilevel converters, often quantified by the output distortion, is critical for optimizing power efficiency, stability, cost, and power density. High fidelity challenges both major modulation categories: carrier-based pulse-width modulation (PWM), and non-PWM approaches, such as nearest-level modulation (NLM). PWM techniques, such as phase-shifted carrier and level-shifted carrier PWM, face a tradeoff between fidelity and switching frequency. Their fidelity is constrained by system efficiency and output bandwidth \cite{10336557}. In contrast, NLM relies on a larger number of modules to match the fidelity of PWM techniques. Hybrid approaches have been explored to combine the strengths of both methods \cite{10336557}. However, a fundamental constraint remains: higher fidelity in multilevel converters ultimately requires more modules to provide additional output levels.

Asymmetric multilevel circuits generate more output levels through the use of different module voltages, which can be combined to form a high number of output voltage levels. Common approaches for creating voltage asymmetry include binary \cite{9771318, 6908214} and tertiary \cite{9070653, 7987947, 5680638} schemes. Our recent research reveals that asymmetric voltage profiles outside these mathematical formulations can also achieve high output quality but also be practical \cite{iecon_ammc_general,10758434}. Whereas the number of output voltage levels of symmetric multilevel converters grows proportionally with the number of modules, this growth is exponential for asymmetric designs.

With a high number of output voltage levels and superior resolution, nearest-level modulation (NLM) 
\cite{sen2022asymmetrically, torres2022selective, jefry202251,chabni2018selective, radhakrishnan2024new, zaid2021symmetric}
is preferred over PWM techniques for MMCs 
\cite{srivastav2018modulation, salman2023design, abdulhakeem2025design, islam2019improvement}
due to its simplicity and easier digital implementation.
In symmetric MMCs, NLM is valued for its low switching rate, as modules typically switch only at the fundamental frequency of the output signal. 
However, these advantages of NLM do not apply to asymmetric multilevel circuits. In asymmetric circuits, NLM is too inflexible, where the switching rate grows exponentially with the number of modules but no longer load all modules similarly.

This paper investigates the non-ideal behavior of asymmetric MMC circuits under NLM. The high number of output levels can cause modules to switch multiple times within a single output period, which leads to transient or even continuous excessive switching of transistors. In extreme cases, gate driver power supplies may be overloaded and drop their output voltage, pushing transistors into saturation and damaging the hardware. Another problem comes from the switching dead time. In bridge circuits, the voltage is not controlled during the dead time but depends on the current direction. In symmetric MMCs, each switching event introduces a step change equal to the voltage of a single module. Consequently, the voltage steps are delayed by the dead time dependent on the current direction, which slightly affects the output and its quality. In asymmetric MMCs, instead, the change at switching events is often mismatched in amplitude compared to the uncontrolled voltage during the dead time --  a minor output adjustment could involve multiple modules and cause a dead-time voltage spike that is significantly larger than the change in the fundamental waveform.

This paper proposes conditional nearest-level modulation (cNLM), a modified version of NLM to solve these problems. We model the challenges NLM faces in asymmetric circuits and incorporate them into cNLM as penalties when determining the nearest level. As a result, cNLM automatically optimizes both output accuracy and switching performance.

\section{Challenges of NLM Application in Asymmetric Multilevel Circuits}
\begin{figure}
    \centering
    \includegraphics{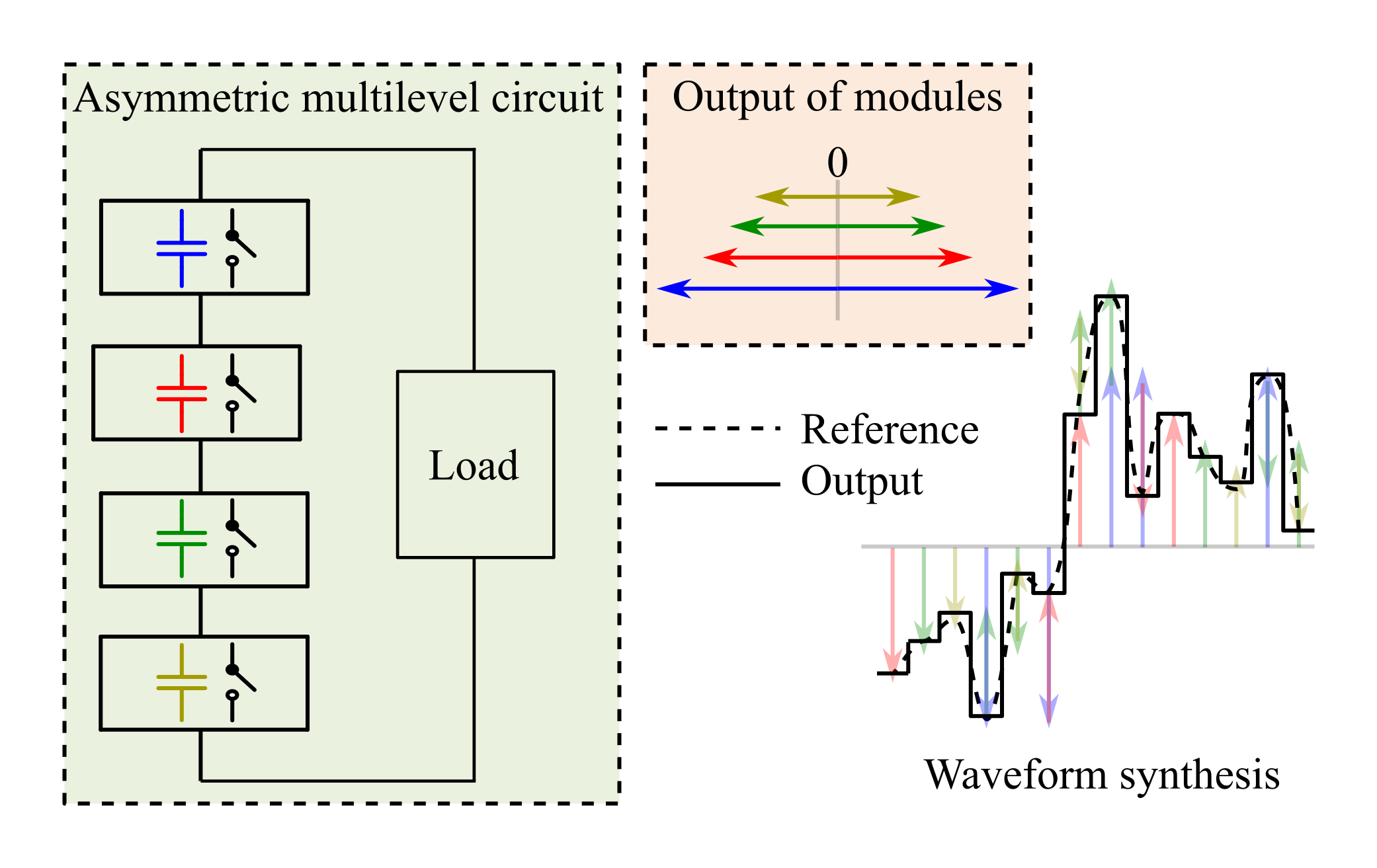}
    \caption{Working principle of asymmetric multilevel converter.}
    \label{fig:illustration_asymmetrical_multilevel}
\end{figure}

In this section, we first outline the working principles of asymmetric multilevel circuits and nearest-level modulation. We then identify two major challenges NLM encounters when applied to asymmetric multilevel circuits. The analysis is performed in the discrete domain for an adaptation to digital control.

\subsection{Generic Model of NLM}
Whether symmetric or asymmetric, all multilevel converters generate the desired output by approximating the continuous reference signals with discrete output levels. Each module can be independently controlled to produce an output of zero, positive, or negative module voltage, $\{0, +1, -1\}$.

Assuming a system consisting of $N$ modules, we can represent its output status with a vector of individual module states per
\begin{equation}
    \overrightarrow{S[k]} = [S_{1}[k],\  S_{2}[k],\ \cdots,\ S_{N}[k]],
\end{equation}
where $S_n[k]$ is the output state of $n^\textrm{th}$ module at moment $k$, and satisfies 
\begin{equation}
    S_n[k] \in \{0, +1, -1\}.
\end{equation}
The output voltage is obtained as 
\begin{equation}
    v_{o}[k] = \overrightarrow{V}\cdot \overrightarrow{S[k]},
\end{equation}
where $\overrightarrow{V}$ is the array of module voltages, per 
\begin{equation}
    \overrightarrow{V} = [V_1, \ V_2, \ \cdots, \ V_N].
\end{equation}

As illustrated in Figure \ref{fig:illustration_asymmetrical_multilevel}, the general goal of modulation is to select the appropriate output vector to minimize the output deviation, per
\begin{equation}
    \overrightarrow{S}[k] = \text{arg }\underset{\overrightarrow{S}}{\text{min }} \Vert v_{\text{ref}}[k] - v_o[k] \Vert,
    \label{equ:generic_model_nlm}
\end{equation}
where the notation $\Vert \cdot \Vert$ represents the deviation between two signals.

The above description applies to both symmetric and asymmetric MMC circuits. The difference between them lies in the uniformity of the voltage distribution. For symmetric circuits, all modules have the same module voltage, per
\begin{equation}
    V_{1} = V_{2} = \cdots = V_{N}, 
\end{equation}
while asymmetric circuits follow
\begin{equation}
    V_{1} \neq V_{2} \neq \cdots \neq V_{N}.
\end{equation}

The asymmetry in module voltages significantly increases the number of output levels. 
The number of available output levels for a symmetric system consisting of $N$ bipolar modules is $2N + 1$, whereas for asymmetric circuits, this number can reach $3^N$. 
This feature provides asymmetric MMCs higher resolution but also introduces additional challenges in modulation.

To obtain the desired output signal with the NLM method, the symmetric MMC simply digitizes the reference signal, i.e., rounds the continuous reference signal into integers, per
\begin{equation}
    v_{o}[k] = \text{round}(v_{\text{ref}}[k]).
\end{equation}
For symmetric MMCs, this process results in low switching rates.

In asymmetric MMC circuits, particularly those that do not follow an integer voltage profile, the efforts to achieve (\ref{equ:generic_model_nlm}) ends up with different switching conditions. Due to a greater quantity of available output levels, asymmetric multilevel converters switch much more frequently than symmetric ones under the NLM scheme.
This difference causes two major issues in asymmetric multilevel converters -- over-switching and spiking.

\subsection{Over-Switching of Modules}
With significantly more output levels than uniform systems, the NLM in asymmetric circuits switches whenever the reference signal passes one of the many levels and switches those modules that form the fine levels notably more. 
Consequently, the switching rates are higher than in symmetric circuits and may even transiently approach the control loop frequency in the worst cases.

Over-switching challenges the hardware. First, the increase in switching rates of transistors raises their losses, which causes overheating and reduces efficiency. Second, over-switching can overload gate drivers and their power supplies, often small isolated dc converters, which can decrease the gate voltage. Consequently, the undervoltage of gate terminal may push transistors into channel saturation and slower switching transitions. 

\subsection{Spikes during Deadtime}
\begin{figure}
    \centering
    \includegraphics{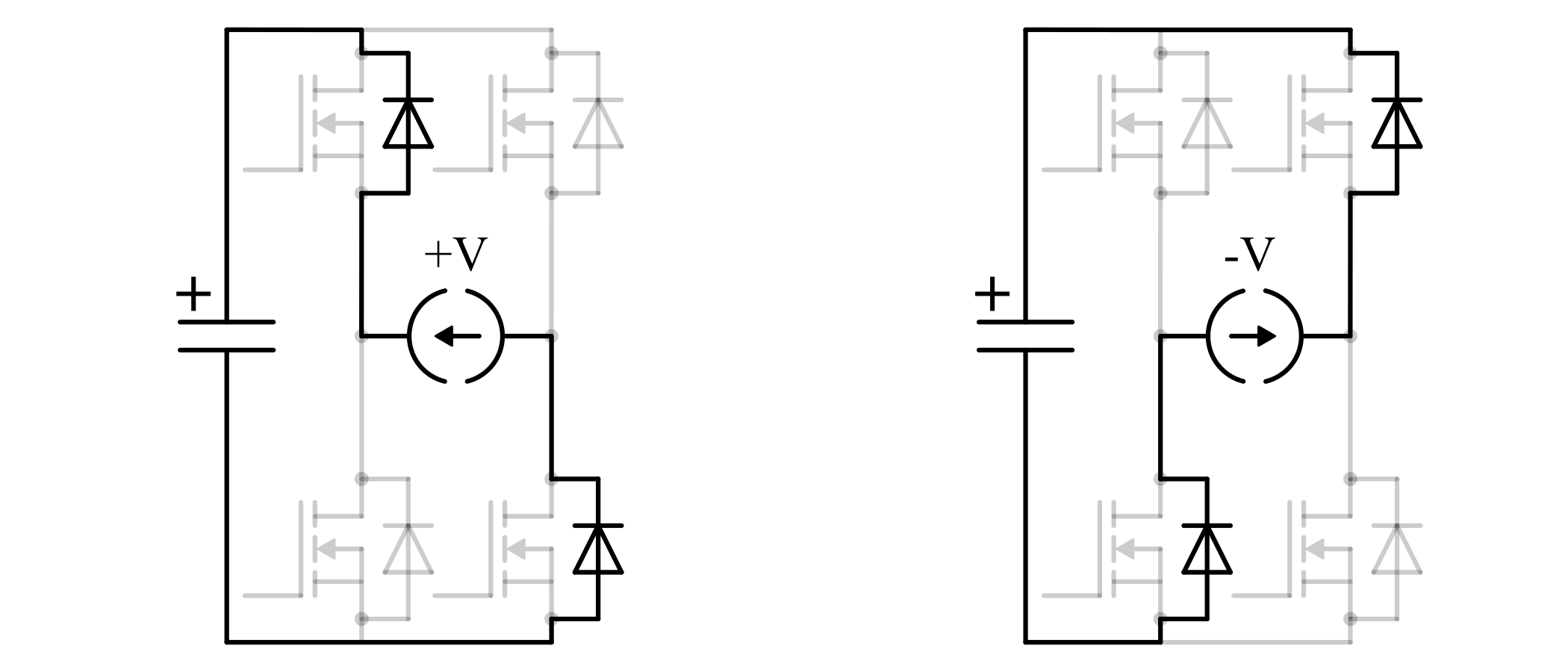}
    \caption{Freewheeling during deadtime.}
    \label{fig:deadtime_freewheeling}
\end{figure}
When switching half bridges from one transistor to the other, an intentional dead time should allow the first transistor to fully turn off before the other one is activated. Dead times typically account for 1\% -- 5\% of the switching period. During the dead time when both high- and low-side transistors are turned off, and the current is freewheeling through separate anti-paralleled or intrinsic body diodes. The output voltage of this module is no longer actively controlled; instead, it depends on the direction of the current as illustrated in Figure \ref{fig:deadtime_freewheeling}. The more modules are involved during the dead time, the greater the load-current-dependent voltage deviation of the output can be. When a minor output change involves multiple modules, the uncontrolled dead-time voltage manifests as spikes. Primary concerns about the spiking phenomenon is the voltage distortion and electromagnetic interference (EMI).

\section{Conditional Nearest-Level Modulation}
We propose the conditional nearest-level modulation (cNLM) to improve switching dynamics in asymmetric multilevel circuits. 

\subsection{Modeling of Challenges}
By definition, the NLM is implemented as
\begin{equation}
\begin{aligned}
    \overrightarrow{S[k]} &= \text{arg }\underset{\overrightarrow{S}}{\text{min }} \left| v_{\text{ref}}[k] - v_o[k] \right| \\
                          &= \text{arg }\underset{\overrightarrow{S}}{\text{min }} \left| v_{\text{ref}}[k] - \overrightarrow{S} \cdot \overrightarrow{V} \right|.
\end{aligned}
\end{equation}
When applying this algorithm directly to asymmetric multilevel circuits, the large number of output levels leads to over-switching and transient spikes. We model these phenomena as follows.

\subsubsection{Overswitching of Individual Modules}
For asymmetric multilevel converters, each module contributes a different voltage, which results in distinct switching sequences. As the switching instances are far from equally spaced out and can lead to bursts, the term switching rate fails to describe the load on individual transistors and their gate drivers. We therefore rate switching  as the interval between consecutive switching events. If a certain module switches at time step $k$, its switching interval is determined as 
\begin{equation}
    \Delta t_{\text{n, sw}}[k] = k - k_{n, \textrm{ls}},
\end{equation}
where $k_{n, \textrm{ls}}$ represents the discrete time step at which the $n^\textrm{th}$ module was last switched. A smaller interval indicates faster switching.
This mechanism can be easily implemented with a single register for each individual module.

\subsubsection{Transient Distortion}
The output voltage of modules during the dead time passively depends on the current direction, as 
\begin{equation}
    v_{n, \text{deadtime}} = -V_{n} \text{sign}(i) D_{n}[k],
\end{equation}
where 
\begin{equation}
    D_{n}[k] = (S_{n}[k-1] \neq S_{n}[k].
\end{equation}
While the output of a module is expected to be $V_{n} S_{n}[k]$, the voltage deviation during the dead time follows  
\begin{equation}
    V_{n,\textrm{d}} = V_{n} (S_{n}[k] + \text{sign}(i) D_{n}[k].
\end{equation}
For the whole system, the deviation due to the dead time can be summarized as 
\begin{equation}
    V_\textrm{d} = \sum_{n=1}^N V_{n} (S_{n}[k] + \text{sign}(i) D_{n}[k]). 
\end{equation}
We can use this result to predict and estimate the voltage spikes during the dead time.

Further, we can simplify it to avoid current sensing or simulation as
\begin{equation}
    V_\textrm{d} = \overrightarrow{V} \cdot \overrightarrow{D[k]}.
\end{equation}

\subsection{Model of Conditional Nearest Level Modulation}
We include these mathematical models as penalty functions to the NLM formulation to form a conditional nearest-level modulation (cNLM) as 
\begin{equation}
    \begin{aligned}
        \overrightarrow{S[k]} &= \text{arg }\underset{\overrightarrow{S}}{\text{min }} \left| v_{\text{ref}}[k] - \overrightarrow{S} \cdot \overrightarrow{V} \right| \\
                              & \quad + \alpha \sum^N_{n=1} D_{n}[k]\frac{1}{\Delta t_{n,\textrm{sw}}} + \beta \overrightarrow{V} \cdot \overrightarrow{D[k]}\\
                              &= \text{arg }\underset{\overrightarrow{S}}{\text{min }}  (O + \alpha P + \beta Q)
    \end{aligned}
    \label{equ:cnlm_equation}
\end{equation}
where
\begin{equation}
\begin{aligned}
    O &= \left| v_{\text{ref}}[k] - \overrightarrow{S} \cdot \overrightarrow{V} \right|,\\
    P &= \sum^N_{n=1} D_{n}[k] \frac{1}{\Delta t_{n, \text{sw}}}, \\
    Q &= \overrightarrow{V} \cdot \overrightarrow{D[k]}. \\
\end{aligned}
\end{equation}

Whereas the nearest-level modulation (represented by part $O$) seeks the switching state that best approximates the reference signal, the conditional NLM (cNLM) balances the impact of switching intervals ($P$) and transient spikes during the dead time ($Q$). 
Through cNLM, we can maintain a high output quality and concurrently manage the switching dynamics. 

\subsection{Variations of cNLM}
From the basic framework of cNLM (\ref{equ:cnlm_equation}), we developed several variations to accommodate constraints for various applications.

\subsubsection{Solid Limit on Switching Intervals}
For applications sensitive to switching, a low limit for the switching interval may take precedence over output quality. Rather than balancing distortion and switching rates, these applications benefit more from a penalty for setting a minimum switching interval, as
\begin{equation}
    P = \sum^N_{n=1} \frac{1}{\text{max}(0, \Delta t_{n, \textrm{sw}} - \Delta T_{sw})},
    \label{equ:variation_solid_limit_tsw}
\end{equation}
where $\Delta T_\textrm{sw}$ represents the minimum allowable interval between consecutive switching events. Any attempt to switch below this limit incurs an infinite penalty and thus is automatically filtered out. This approach effectively avoids transient switching stress for transistors. 

\subsubsection{Module-Specific Over-Switching Penalty}
Since asymmetric modules operate at different voltages, the key components---transistors, capacitors, and auxiliary circuits---can vary across modules. These differences result in varying switching performance and limits. Therefore, we designed module-specific switching penalty factors and refined the over-switching penalty, i.e., $\alpha P$, to 
\begin{equation}
    \alpha P = \sum^N_{n=1} \alpha_{n} D_{n}[k]\frac{1}{\Delta t_{n, \textrm{sw}}}.
    \label{equ:variation_different_alpha}
\end{equation}
By assigning different values to $\alpha_{n}$ for each module, we can fine-tune the penalty degree according to their individual switching constraints. 

\subsubsection{Precise Form of Spiking Penalty}
The proposed model (\ref{equ:cnlm_equation}) simplified the spiking penalty to eliminate the need for current sampling and reduce the computational load. However, for systems with access to current sensor signals or observers, 
the precise form of $Q$ follows
\begin{equation}
    Q = \sum_{n=1}^N V_{n} (S_{n}[k] + \text{sign}(i) D_{n}[k].
    \label{equ:variation_precise_spike}
\end{equation}

\subsubsection{Mathematical Variations}
The components of cNLM can use different norms. For instance, the term $O$ can use a squared or higher-order norm 
\begin{equation}
    O = (v_{\text{ref}}[k] - \overrightarrow{S} \cdot \overrightarrow{V})^2.
\end{equation}
Similarly, we can also generalize $P$ to 
\begin{equation}
    P = \sum^N_{n=1} \left(D_{n}[k]\frac{1}{\Delta t_\textrm{sw}}\right)^p,
\end{equation}
where $p$ may be adjusted to have different penalty curve of switching intervals.

\section{Results}
\begin{figure}[H]
    \centering
    \includegraphics{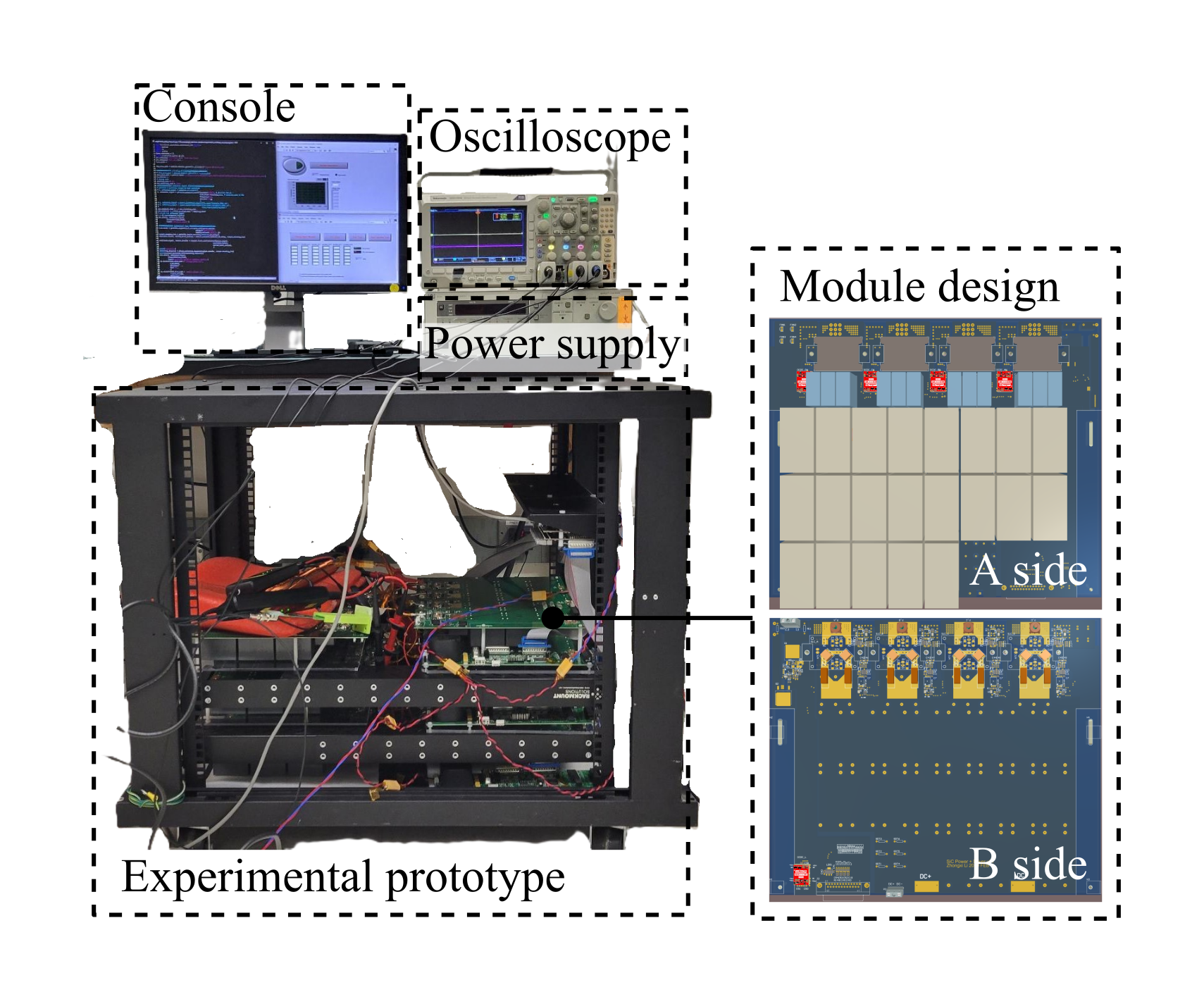}
    \caption{The four-module prototype and test platform.}
    \label{fig:photo_prototype}
\end{figure}
\begin{figure}[H]
    \centering
    \begin{subfigure}[b]{0.48\textwidth}
        \centering
        \includegraphics[width=\linewidth]{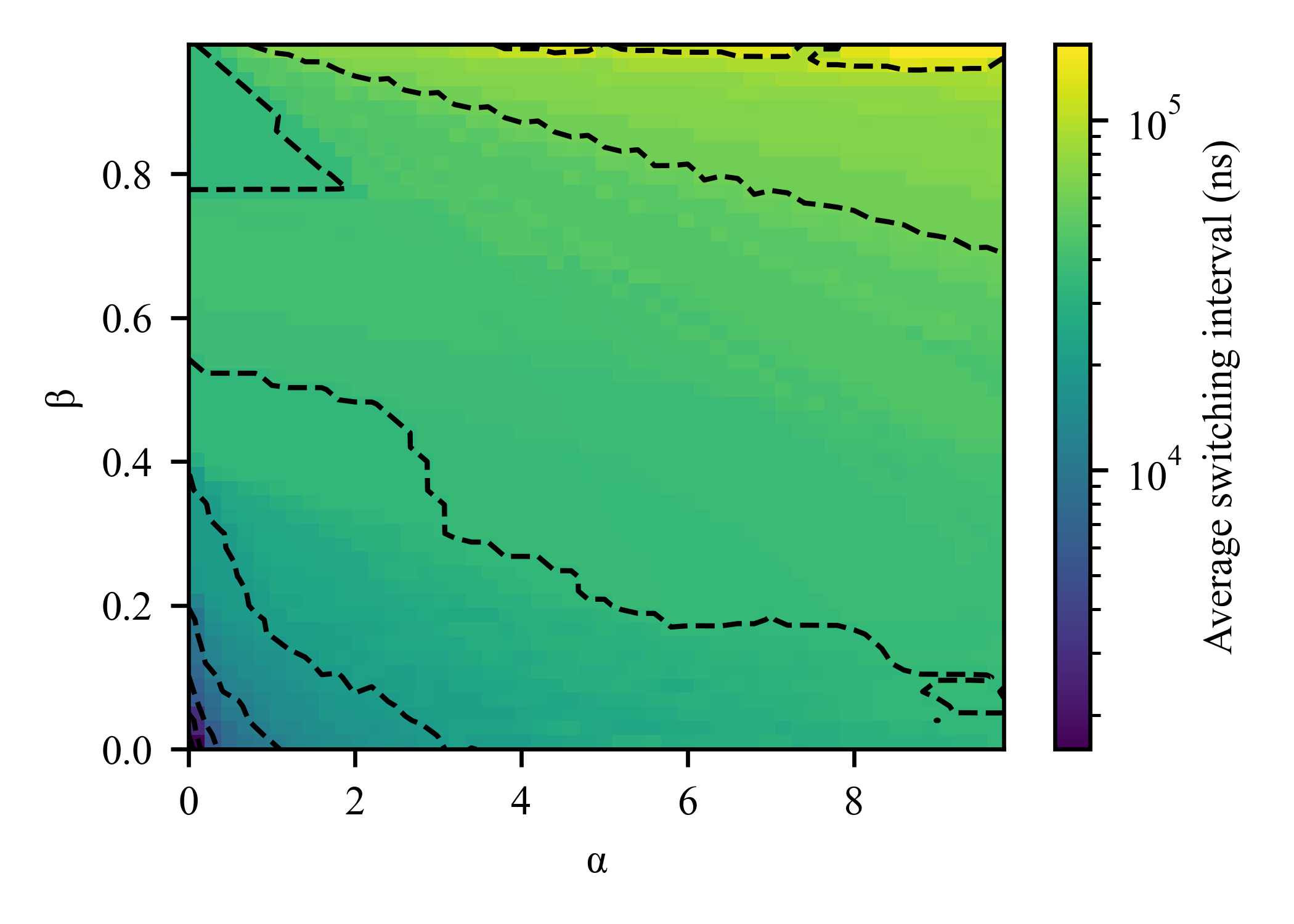}
        \caption{Influence on switching density.}
        \label{fig:influence_on_switching_rate_alpha_beta}
    \end{subfigure}
    \hfill
    \begin{subfigure}[b]{0.48\textwidth}
        \centering
        \includegraphics[width=\linewidth]{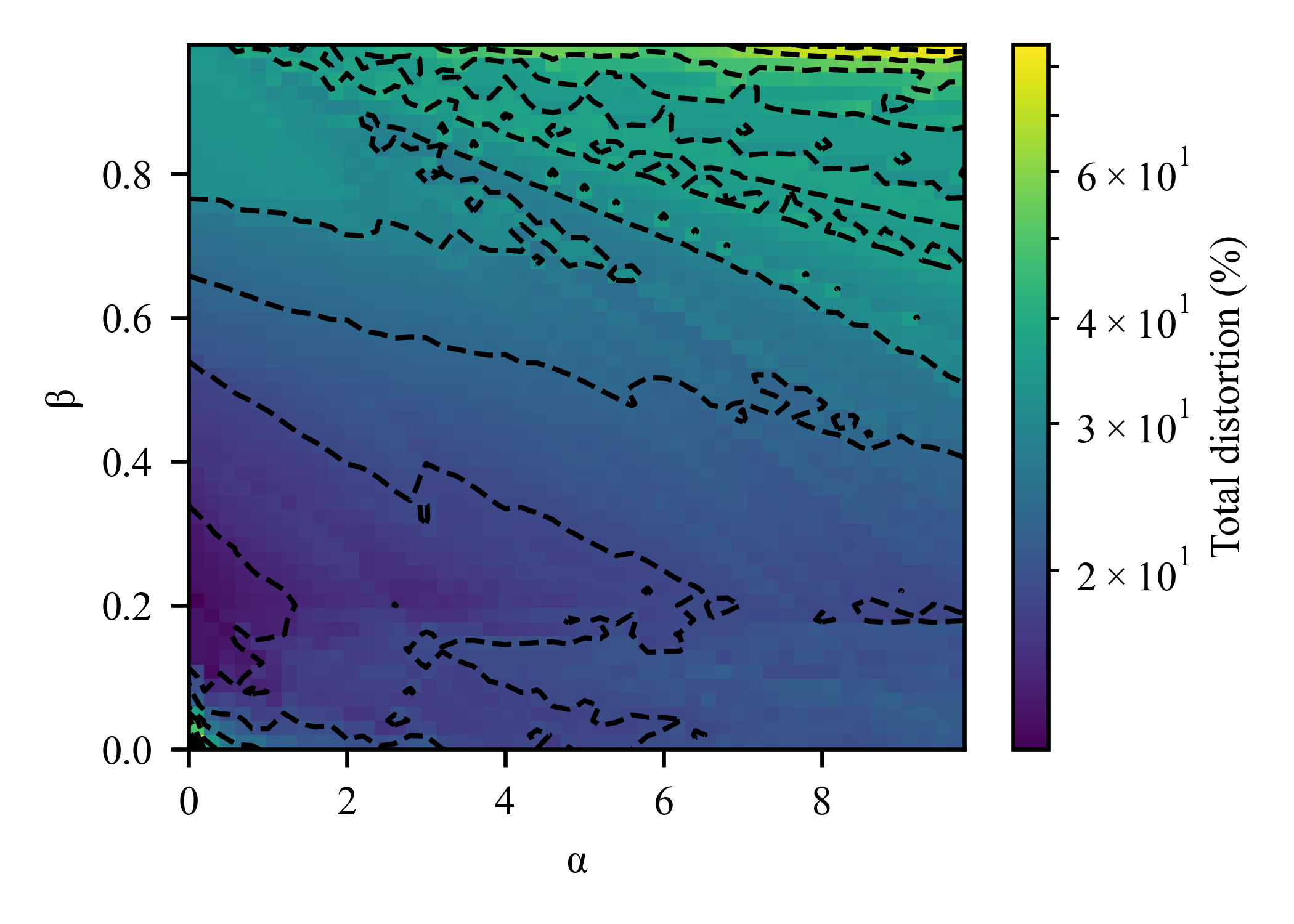}
        \caption{Influence on output quality.}
        \label{fig:influence_on_td_alpha_beta}
    \end{subfigure}
    \caption{Exploration of the influence of penalty factors.}
    \label{fig:combined_influence_alpha_beta}
\end{figure}
\begin{figure*}[h]
    \centering
    \includegraphics[width=0.95\textwidth]{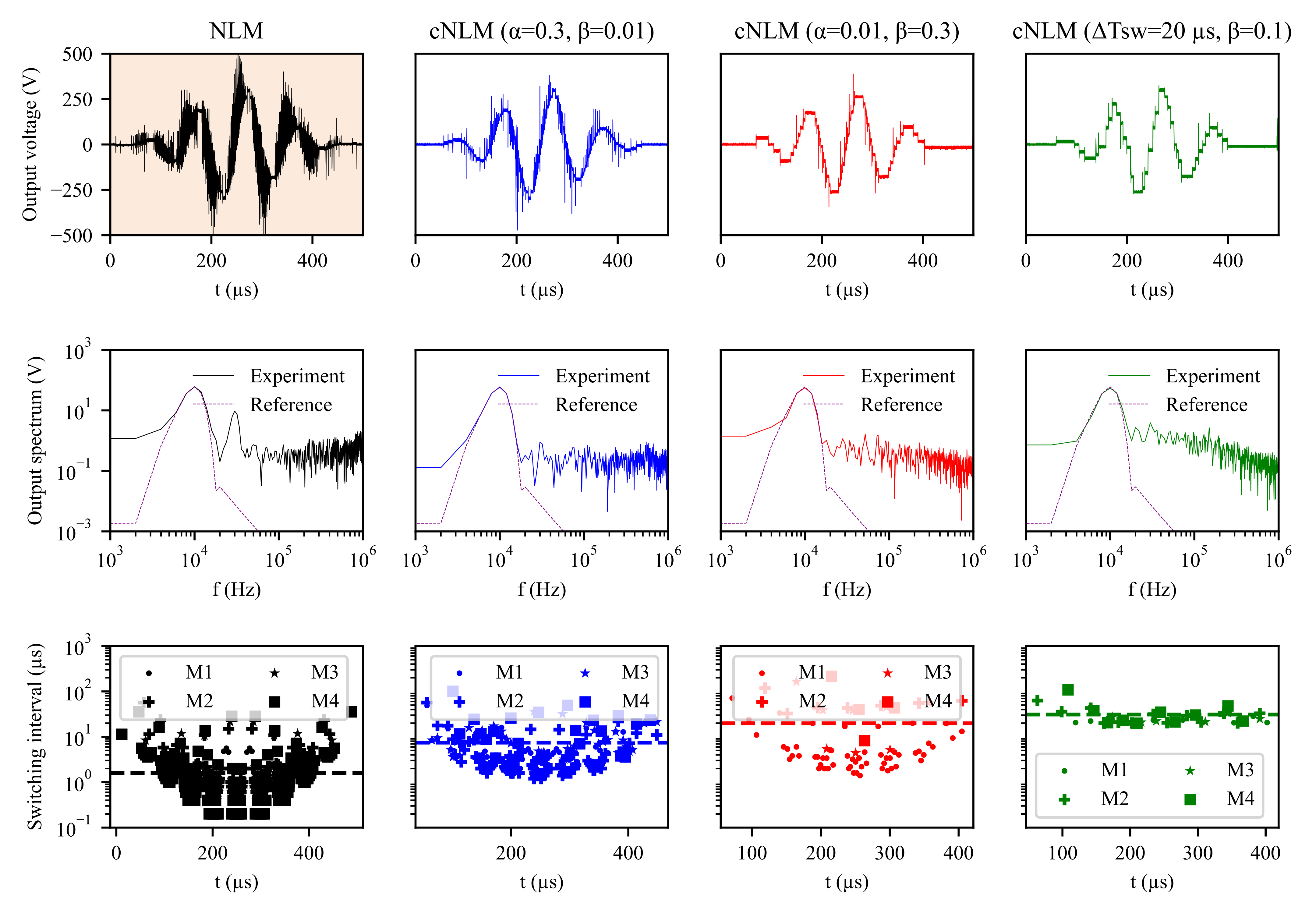}
    \caption{Comparison of experimental performance between different parameters and variations of conditional NLM.}
    \label{fig:comparison_of_different_cnlm_results}
\end{figure*}

\begin{table*}
    \caption{Switching Behavior and Output Performance of Different Modulation Methods}
    \label{tab:modulation_comparison}
    \centering
    \begin{tabular}{l|c|c|c}
        \hline
        \textbf{Modulation Method} & \textbf{Minimum Switching Interval} & \textbf{Average Switching Rate} & \textbf{Voltage Total Distortion} \\
        \hline
        NLM & 200~ns & 627.0~kHz & 66.3\,\% \\
        cNLM ($\alpha = 0.3, \beta = 0.02$) & 1,200~ns & 134.1~kHz & 17.4\,\% \\
        cNLM ($\alpha = 0.01, \beta = 0.3$) & 1,400~ns & 50.2~kHz & 15.1\,\% \\
        cNLM (min switching at 20 µs, $\beta = 0.1$) & 20,000~ns & 32.2~kHz & 22.2\,\% \\
        \hline
    \end{tabular}
\end{table*}

\begin{figure*}
    \centering
    \includegraphics{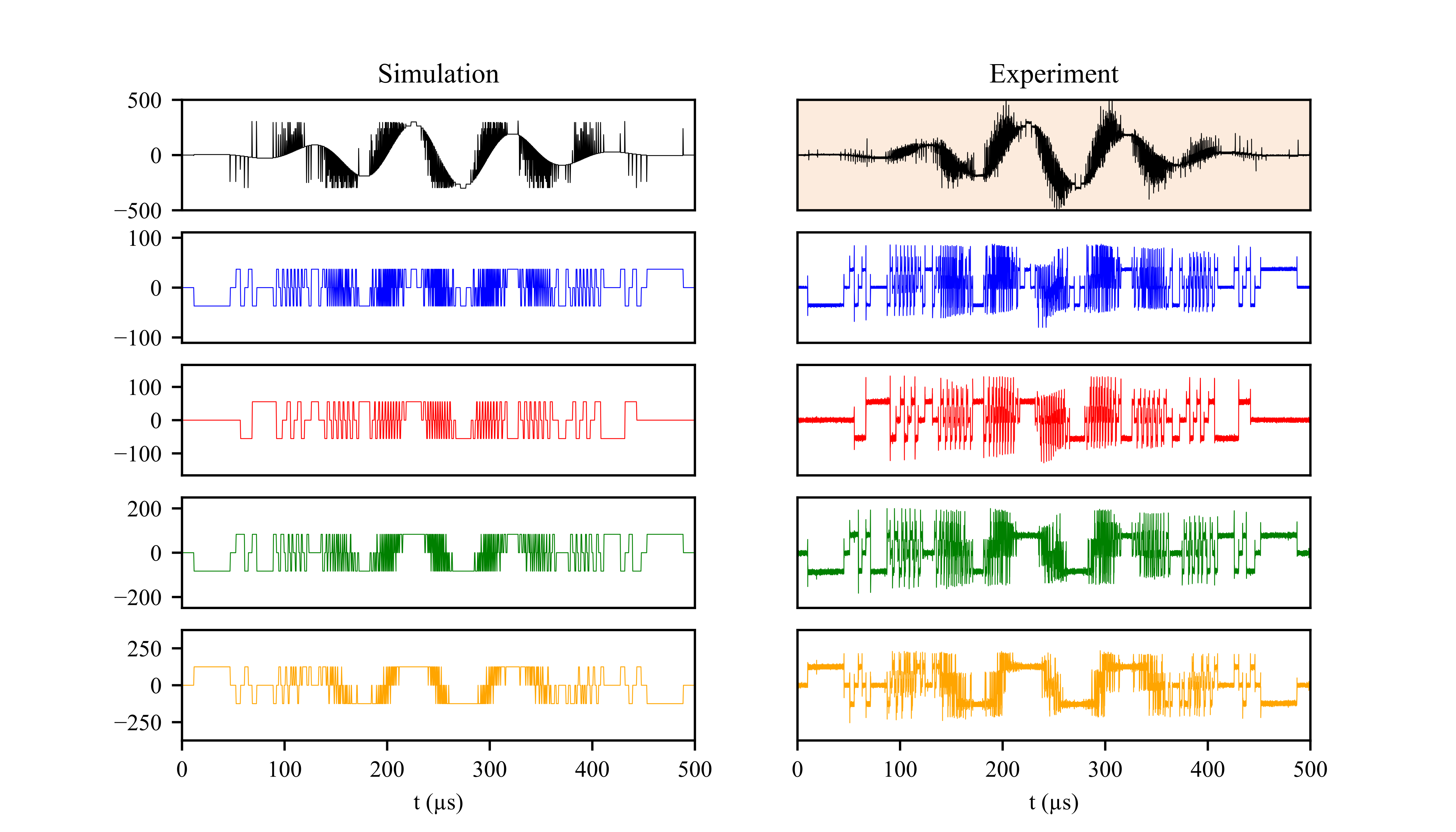}
    \caption{Validation of predicting switching noise and over-switching issues in asymmetric multilevel converters under nearest-level modulation.}
    \label{fig:verification_of_switching_issues}
\end{figure*}

\subsection{Experimental Platform and Test Conditions}
We constructed a four-module prototype with field-effect transistors (FET) and isolated gate drivers (Figure \ref{fig:photo_prototype}) \cite{zhang2025asymmetric}. This setup supports a continuous switching rate of up to 50 kHz, with the minimum switching interval, constrained by the dead-time length (approximately 500 ns). The prototype drives an inductive load of 14 µH. Although our findings and solutions are applicable to commonly used binary and ternary configurations, we chose a geometric voltage pattern with a ratio of 1.5 to reduce the voltage gap for better practicality \cite{iecon_ammc_general}. For a system output voltage of 300~V, four modules are charged to 37~V, 55~V, 83~V, and 125~V through a programmable voltage supply (HP 6030A). Results are collected with a wide-bandwidth oscilloscope (MDO3054, 2.5 GSa/s, 4 channel, Tektronix Co.).

We adopt a Gaussian polyphasic reference signal with a fundamental frequency of 10 kHz and a standard deviation of $8 \times 10^{-5}$. 
This signal serves as a benchmark for different modulation approaches, as it covers a full range of modulation indices in the time domain and a broad spectrum in the frequency domain.

\subsection{Influence of Penalty Factors}
Prior to implementing cNLM algorithms, we analyzed the influence of penalty factors $\alpha$ and $\beta$ in  a large-scale simulation. Figure \ref{fig:influence_on_switching_rate_alpha_beta} visualizes their impact on switching intervals,  Figure \ref{fig:influence_on_td_alpha_beta} in turn the output fidelity. 

The penalties tend to make the modulation more conservative. An increase in either parameter accordingly resulted in longer switching intervals, which in turn reduced the overall switching rate (Figure \ref{fig:influence_on_switching_rate_alpha_beta}).

The influence on the total distortion of the output voltage followed a different trend. As shown in Figure \ref{fig:influence_on_td_alpha_beta}, the penalties suppressed the switching spikes, which initially improved the output quality. However, as the penalty factors increased further, the total distortion began to rise due to the fundamental misalignment in the waveform. 

\subsection{Performance Improvements by cNLM}
We proceed with three cNLM configurations for further evaluation, including two configurations with different parameter sets, and one variation with a limit on the minimum switching interval. We also performed a traditional NLM trial as a control.

Figure \ref{fig:comparison_of_different_cnlm_results} summarizes their performance through their output waveform, spectrum, as well as switching density. Their spectra are compared to the reference Gaussian signal, as shown in the second row. Their switching events are visualized with a scatter plot, where each point represents a single switching event; different scatter shapes correspond to different modules. The x-value of the scatter marks the time at which the switching occurs, whereas the y-axis indicates the interval since the previous switching event. The average switching interval is marked by a dashed horizontal line. Table \ref{tab:modulation_comparison} quantifies these results.

Compared to the original NLM method, all cNLM trials significantly reduced both the switching rate and the occurrence of spikes.
Specifically, large values of $\alpha$ and small values of $\beta$, such as the first cNLM trial ($\alpha = 0.3$, $\beta = 0.01$, second column in Figure \ref{fig:comparison_of_different_cnlm_results}), offer a high output resolution, while they reduce the average switching rate to only one fifth of the original NLM. 

On the other hand, the trial with a larger value of $\beta$ and smaller values of $\alpha$ dominates the conditional modulation with the spiking penalty. This trial further reduces the occurrence of spikes. 
As shown in the third column of Figure \ref{fig:comparison_of_different_cnlm_results}, this configuration results in a total distortion of 15.1\,\%, down from 66.3\,\% as in the original NLM. 
Meanwhile, its average switching rate is reduced to only 8\,\% of the original NLM. 

The third cNLM trial (fourth column in Figure \ref{fig:comparison_of_different_cnlm_results}) aims to prevent temporary overload of transistors through a limit on the minimum switching interval at 20~µs. While the first two cNLM trials significantly reduced the average switching rate, their minimum switching intervals still reached as low as 1200~ns and 1400~ns. By contrast, this variation clipped the minimum switching interval at 20~µs. The trade-off of this approach is an increase in output distortion, as the constraint on switching interval may deny timely switching and cause delays in the output.

\subsection{Transistor Over-Switching Problems}
We zoomed in the result of NLM to validate our assumption of transistor over-switching. The experimental results of NLM, including both system and individual module output, align with our theoretical predictions. All modules have experienced a certain level of over-switching, albeit unevenly. The minimum switching interval for a single module in NLM drops to 200~ns, while the first module exhibited a denser switching pattern than the others. This phenomenon indicates substantial switching stress on the transistors. 

By contrast, the cNLM algorithms mitigated this risk. The minimum switching intervals with the three cNLM configurations are respectively prolonged to 1200 ns, 1400 ns, 20000 ns. Particularly, the configuration of a solid switching interval floor can flexibly adjust this value.

\section{Conclusion}
This paper presented a new modulation method---conditional nearest level modulation (cNLM)---to improve the switching dynamics in asymmetric multilevel converters. By incorporating additional penalties into the nearest-level calculation, the proposed cNLM algorithm efficiently reduces both switching rates and the occurrence of output spikes. Additionally, multiple variations are presented to serve special functions, such as enforcing a minimum switching interval. 

\bibliographystyle{Bibliography/IEEEtranTIE}
\bibliography{Bibliography/IEEEabrv,Bibliography/main}\ 

\end{document}